\documentclass[aip,pof,reprint,a4paper,floatfix]{revtex4-1}
\usepackage{graphicx}
\usepackage{bm}
\usepackage{dcolumn}
\usepackage{amsmath}
\usepackage{color}
\usepackage[normalem]{ulem}   
\usepackage[bookmarks=true,colorlinks=true,linkcolor=blue,citecolor=blue]{hyperref}
\newcommand{\ben} {\begin{equation}}
\newcommand{\een} {\end{equation}}
\newcommand{\be} [1] {\begin{equation} \label{#1}}
\newcommand{\ee} {\end{equation}}
\newcommand{\bse} [1] {\begin{subequations} \label{#1}}
\newcommand{\ese} {\end{subequations}}
\newcommand{\ban} {\begin{eqnarray*} }
\newcommand{\ean} {\end{eqnarray*} }
\newcommand{\bea} {\begin{eqnarray}}
\newcommand{\eea} {\end{eqnarray}}

\newcommand{\dashed}{\hbox{{--}\,{--}\,{--}\,{--}}}

\newsavebox{\mybox}
\sbox{\mybox}{\dashed}

\begin{document}
\title{%
Travelling-waves consistent with
turbulence-driven secondary flow in a square duct} 
\author{
Markus Uhlmann$^1$, Genta Kawahara$^2$ and Alfredo Pinelli$^3$\\
{\small
  $^1$Institute for Hydromechanics, 
  Karlsruhe Institute of Technology,
  76128 Karlsruhe, Germany}\\
{\small
  $^2$Department of Mechanical Science, Osaka University, 560-8531 Osaka,
  Japan}\\
{\small
  $^3$Modeling and Numerical Simulation Unit, CIEMAT, 28040
  Madrid, Spain}
}
\date{\today}
\begin{abstract}
  We present numerically determined travelling-wave solutions for 
  pressure-driven flow through a straight duct with a square
  cross-section.  
  This family of solutions represents typical coherent structures (a
  staggered array of counter-rotating streamwise vortices and an
  associated low-speed streak) on each wall.
  Their streamwise average flow in the cross-sectional plane corresponds
  to an eight vortex pattern much alike the secondary flow found in the
  turbulent regime.  
\end{abstract}
\maketitle
\section{Introduction}\label{sec-intro}
The flow through a straight duct with rectangular cross-section shares
important features with both circular pipe flow and plane channel
flow. Analogous to pipe flow, laminar flow in a {\it square} duct is
believed to be stable with respect to all infinitesimal perturbations
\cite{tatsumi:90}. 
On the other hand, for aspect ratios above a critical value of
$A_{crit}\approx3.2$, duct flow is linearly unstable.
In fact, ducts with large aspect ratio are often used in laboratory
experiments as a substitute for plane channels\cite{nishioka:85}.
Despite these similarities, the fact that the cross-section of
rectangular ducts is neither axisymmetric nor spanwise-homogeneous
leads to an interesting phenomenon: 
{\it turbulent} flow in this geometry 
exhibits mean secondary motion in the plane perpendicular to its axis
\cite{nikuradse:26}. 
This mean cross-flow motion is directed towards the
corners in the vicinity of the diagonals, and away from the wall near
the bisectors, giving rise to a pattern with eight vortices. 
Although of small amplitude (few percent of the primary flow), the
secondary motion significantly deforms the latter, thereby
inducing strong variations of local mean wall-shear stress along the
edges \cite{gavrilakis:92}.

The results of recent direct numerical simulations of turbulent square
duct flow suggest that the occurrence of secondary motion is a
statistical footprint of the preferential location of coherent
structures at certain positions along the edges
\cite{uhlmann:07a}. 
This conclusion is based upon data obtained 
through a vortex eduction study applied to snapshots of turbulent flow
fields at low Reynolds numbers. 
In contrast to these ``approximate'' coherent structures, ``exact''
coherent structures (travelling-waves,
time-periodic solutions) 
in duct flow have only received little attention in the literature.  
To our knowledge, the only study presenting non-linear travelling-wave
solutions for the square duct is ref.~\citenum{wedin:09}.
However, in contrast to the claim of the authors, the solution
family presented therein is not representative of the turbulent flow
states observed in ref.~\citenum{uhlmann:07a}. 
More specifically, the travelling-waves of ref.~\citenum{wedin:09}
exhibit a pattern of four mean secondary flow vortices which can
appear in two orientations (rotated by $\pi/2$).
In this case what should be compared is on the one hand the
  statistically averaged turbulent flow field and on the other hand
  the average between the two (equivalent) travelling-wave solutions
  rotated by $\pi/2$ with respect to each other.
Performing this averaging on the solution of ref.~\citenum{wedin:09},
however, results in a vorticity pattern with a sign which is opposite
to the one of the mean secondary flow found in
turbulence\cite{uhlmann:07a}.   

The objective of the present work is to determine exact coherent
structures which are potentially relevant to turbulent
flow in a square duct.  
For this purpose we have computed different families of
travelling-waves.  
In this article 
we present one particular solution family which entails
an eight-vortex secondary flow with the same symmetries and sense of 
rotation as that found in time-averaged turbulent fields. 
This enables us to directly compare the properties of the exact
coherent structures to statistical quantities measured in turbulent
flow. We are therefore in a position to give solid theoretical
support to our earlier conjecture\cite{uhlmann:07a,pinelli:09a} by
showing that near-wall coherent structures are indeed capable of
generating secondary flow in a square duct. 
%
\begin{table}
  \begin{center}
  \begin{minipage}{.4\linewidth}
    \centerline{$(a)$}
    \begin{tabular}{lll}
      \multicolumn{1}{c}{$N_x$}&
      \multicolumn{1}{c}{$c_f$}&
      \multicolumn{1}{c}{${\cal E}$}
      \\\hline
      2&0.0361062&8.5279e-3\\
      4&0.0364307&3.8270e-4\\
      6&0.0364184&4.5376e-5\\
      8&0.0364168&1.4451e-6\\
      10&0.0364167&5.6946e-8\\
      14&0.0364167&---
    \end{tabular}
    \end{minipage}
    \\[1ex]
  \begin{minipage}{.4\linewidth}
    \centerline{$(b)$}
    \begin{tabular}{cll}
      \multicolumn{1}{c}{$N_y=N_z$}&
      \multicolumn{1}{c}{$c_f$}&
      \multicolumn{1}{c}{${\cal E}$}
      \\\hline
      20&0.0320227&0.00526353\\
      30&0.0321351&0.00176935\\
      34&0.0321557&0.00113064\\
      40&0.0321759&0.00050299\\
      44&0.0321851&0.00021773\\
      48&0.0321921&---
    \end{tabular}
    \end{minipage}
    \end{center}
  \caption{Convergence of the solution with the truncation level. 
    $(a)$ Varying the number of Fourier modes for fixed $N_y=N_z=30$
    ($Re=2345$, $\alpha h=1$, lower branch).
    $(b)$ Varying the number of modified Chebyshev polynomials for
    fixed $N_x=2$ 
    ($Re=2600$, $\alpha h=1.5547$, lower branch).
    The relative error is computed with respect to the result obtained
    at the highest truncation level $c_{f,ref}$, viz.\ 
    ${\cal E}\equiv(c_f-c_{f,ref})/c_{f,ref}$.
  }
  \label{tab-convergence}
\end{table}
\section{Numerical method}\label{sec-numa}
In the following the coordinates $\mathbf{x}=(x,y,z)$ are chosen such
that $x$ is aligned with the duct axis and $y,z$ are oriented along
the duct bisectors.  
%
From the duct half-width $h$, the maximum laminar flow velocity
$u_{max}$ and the kinematic viscosity $\nu$ we can form the base flow
Reynolds number $Re=u_{max}h/\nu$. 
We will also use the bulk flow Reynolds number $Re_b=u_bh/\nu$ (based
on the bulk velocity $u_b$) as well as the friction-velocity
based Reynolds number $Re_\tau=u_\tau h/\nu$ (where the wall-friction
velocity $u_\tau$  is defined in the usual way). The superscript $+$
will be used to represent quantities normalized with $\nu$ and
$u_\tau$. 
Introducing the decomposition of the velocity field 
$\mathbf{u}=U({y},z)\mathbf{e}_x+\mathbf{u}^\prime(\mathbf{x},t)$ and the
pressure ${p}=P(x)+{p}^\prime(\mathbf{x},t)$, the flow equations for the
perturbations read: 
\begin{equation}
  \label{equ-ns}
  \begin{array}{l}
    \partial_t\mathbf{u}^\prime
    +U\partial_x \mathbf{u}^\prime
    +(\mathbf{u}^\prime\cdot\nabla)(U\mathbf{e}_x+\mathbf{u}^\prime)
    +\nabla p^\prime
    =\nu\nabla^2\mathbf{u}^\prime,\\
    \nabla\cdot\mathbf{u}^\prime=0,\quad
    \mathbf{u}^\prime(x,y=\pm h,z,t)=\mathbf{u}^\prime(x,y,z=\pm h,t)=0,
  \end{array}
\end{equation}
where $\mathbf{e}_x$ is the unit vector in the $x$-direction,
$\mathbf{u}^\prime=(u^\prime,v^\prime,w^\prime)$, the mass density
has been taken as unity and $\partial_t=\partial/\partial t$ as well
as $\partial_x=\partial/\partial x$. 
The streamwise laminar base flow component $U$ is solution of
$(\partial^2/\partial y^2+\partial^2/\partial z^2)U=
Re\cdot\mbox{d}P/\mbox{d}x$ 
with $\mbox{d}P/\mbox{d}x=const$.  
We employ a primitive variable formulation, expanding 
the dependent variables
$\varphi=\{u^\prime,v^\prime,w^\prime,p^\prime\}$ as follows 
(assuming the solution to be a travelling-wave): 
\begin{equation}\label{equ-1}
  \begin{split}
    \displaystyle
    \varphi(\mathbf{x},t)=
    \sum_{n=-N_x}^{N_x}\sum_{m=0}^{2N_y-k^{(\varphi)}}
    \sum_{l=0}^{2N_z-k^{(\varphi)}}
    \varphi_{n\,m\,l}\,\times\\
    \times\,\phi_m^{(\varphi)}(y)\,\phi_l^{(\varphi)}(z)
    \,\exp(\mbox{i}n\alpha(x-ct))\,,
  \end{split}
\end{equation}
where $\alpha$ is the streamwise wavenumber, $c$ the (real-valued) 
phase speed and $\mbox{i}=\sqrt{-1}$. The functions $\phi^{(\varphi)}$ are
modified Chebyshev polynomials which---in the case of $\varphi$ being
a velocity component---incorporate 
the wall boundary conditions \cite{uhlmann:05b}. 
No boundary conditions are prescribed for the pressure field, and
hence two polynomial degrees less than for the velocity
field are used, i.e.\ $k^{(p^\prime)}=2$ and
$k^{(u^\prime)}=k^{(v^\prime)}=k^{(w^\prime)}=0$.  
The equations (\ref{equ-ns}) admit solutions which exhibit particular
combinations of odd/even cross-sectional parities as described
below. By restricting our attention to these symmetric solutions, the
number of modes in the $y$ and $z$ directions can be halved without
loss of resolution.  
%
\begin{figure}
  \begin{center}
    \begin{minipage}{2ex}
      $(a)$\\[5ex]
      $\displaystyle\frac{E_{3D}^{1/2}}{u_b}$
    \end{minipage}
    \begin{minipage}{.75\linewidth}
      \includegraphics[width=\linewidth]{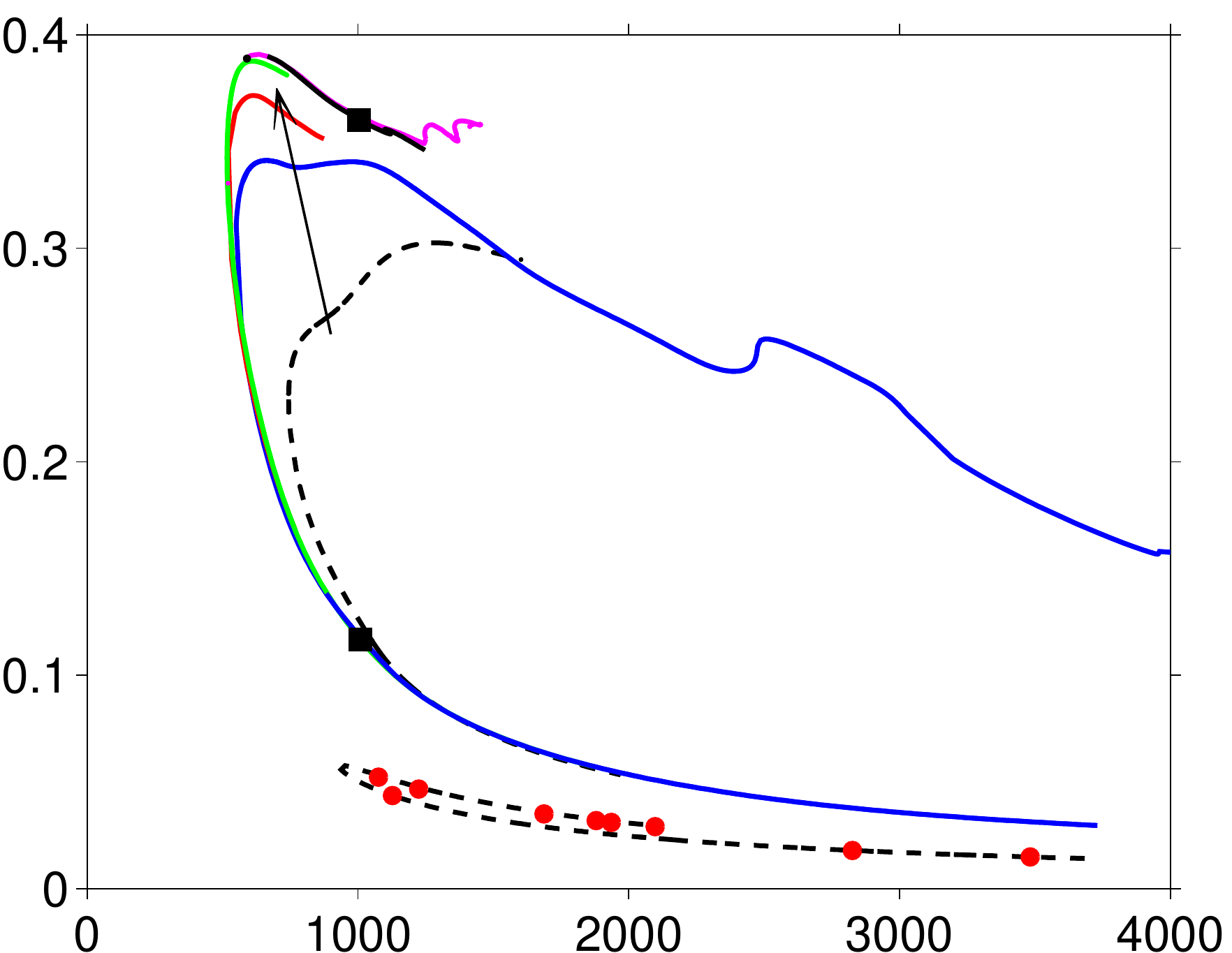}
      \hspace*{-.55\linewidth}\raisebox{.6\linewidth}{
        $\alpha h=0.6$
      }
      \hspace*{-.65\linewidth}\raisebox{.08\linewidth}{
        $1.5547$
      }
      \hspace*{+.5\linewidth}{}
      \\
      \centerline{$Re_b$}
    \end{minipage}
    \\
    \begin{minipage}{2ex}
      $(b)$\\[5ex]
      $\displaystyle\frac{E_{3D}^{1/2}}{u_b}$
    \end{minipage}
    \begin{minipage}{.75\linewidth}
      \includegraphics[width=\linewidth]{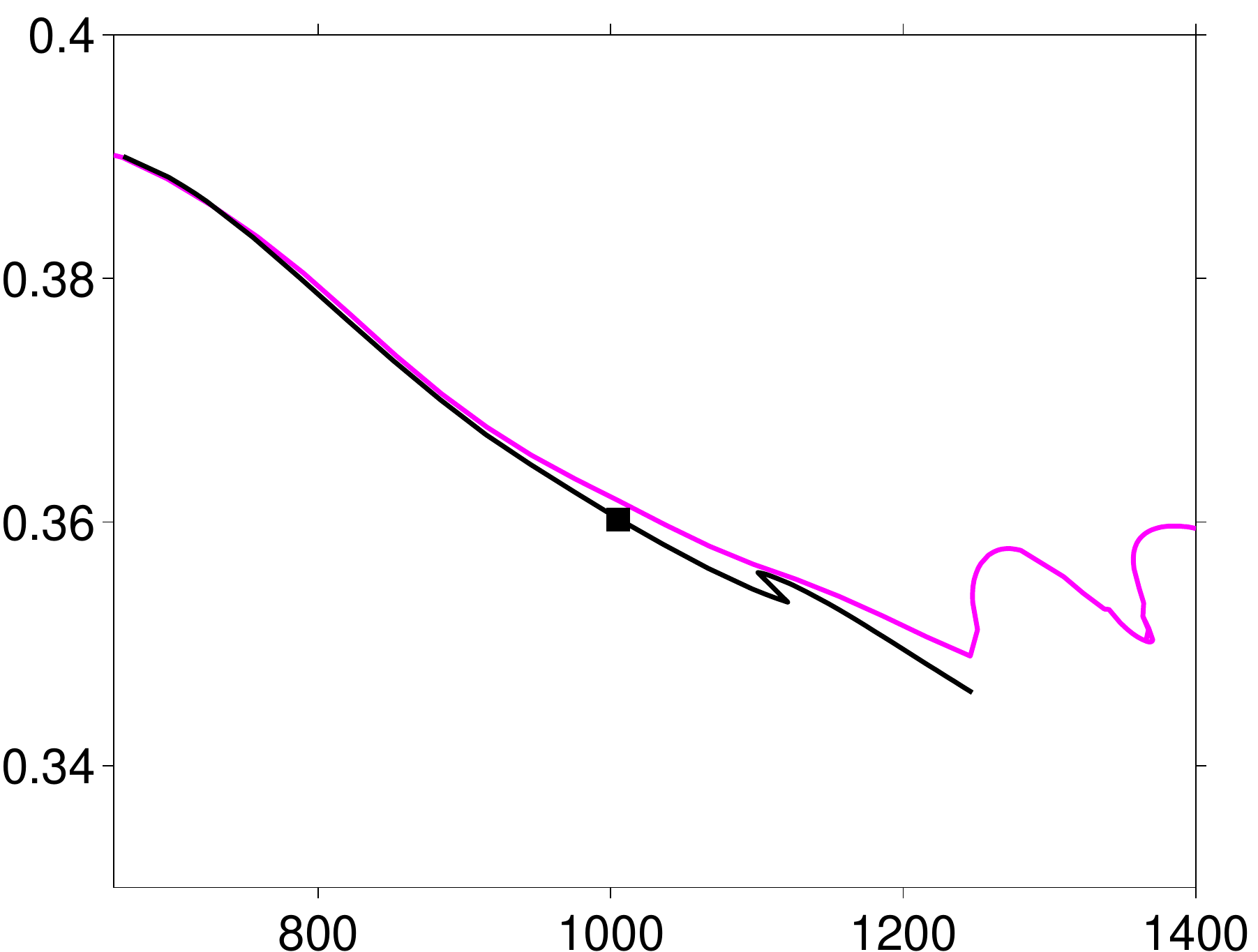}
      \hspace*{-.5\linewidth}\raisebox{.35\linewidth}{
        $14\times20\times20$
      }
      \hspace*{-.56\linewidth}\raisebox{.2\linewidth}{
        $20\times20\times20$
      }
      \hspace*{+.15\linewidth}{}
      \\
      \centerline{$Re_b$}
    \end{minipage}
  \end{center}
  \caption{%
    Convergence of the solution curves for two different streamwise
    wavenumbers $\alpha h=\{0.6,1.5547\}$ computed at different
    truncation levels. 
    $(a)$ 
    For $\alpha h=1.5547$, the dashed line
    corresponds to $2\times30\times30$ and the filled circles to
    $4\times32\times32$. 
    For $\alpha h=0.6$, the truncation level varies along the arrow in
    the order: $2\times30\times30$, $4\times20\times20$,
    $6\times26\times26$, $10\times20\times20$, $14\times20\times20$,
    $20\times20\times20$, with the filled squares indicating
    individual solutions at $20\times26\times26$.
    $(b)$ Close-up of the data shown in $(a)$ for $\alpha h=0.6$ on the
    upper branch. The curve for the truncation level
    $20\times20\times20$ is drawn up to the point where the relative
    difference with respect to the solution obtained at
    $14\times20\times20$ is still less than one percent. 
  }
  \label{fig-cf-conv1}
\end{figure}
\begin{table}
  \begin{center}
    \begin{tabular}{cllll}
      &\multicolumn{1}{c}{$u$}&
      \multicolumn{1}{c}{$v$}&
      \multicolumn{1}{c}{$w$}&
      \multicolumn{1}{c}{$p$}\\
      I&(o,e)&(e,e)&(o,o)&(o,e)\\ 
      II&(o,o)&(e,o)&(o,e)&(o,o)\\
      III&(e,e)&(o,e)&(e,o)&(e,e)\\
      IV&(e,o)&(o,o)&(e,e)&(e,o)
    \end{tabular}
  \end{center}
  \caption{The four combinations of cross-sectional solution parities
    admitted by the linearized operator, as shown in
    ref.~\citenum{tatsumi:90}. The notation is such that e.g.\ ``(o,e)''
    stands for an odd parity in $y$ and an even parity in $z$.}
  \label{tab-parities}
\end{table}
After substituting (\ref{equ-1}) into (\ref{equ-ns}), a Galerkin
method is employed in the streamwise (Fourier) direction, and a 
collocation method is used in the cross-stream (Chebyshev)
directions.  
In our case, the collocation points for pressure are
chosen as the {\it Gauss} points, i.e.\ pressure is staggered
with respect to the usual {\it Gauss-Lobatto} grid used for velocity, 
thereby avoiding a discretization of the continuity equation at the
corner points. 
It should be noted that the above discretization of the space operator
guarantees a solution which is free from spurious pressure
modes\cite{canuto:07}.  
The solution of the resulting non-linear algebraic system is performed
via Newton-Raphson iteration, the linear system being solved by means
of LU decomposition, exploiting parallel computing techniques.
Continuation along the different problem parameters is
achieved through an arc-length procedure \cite{rheinboldt:83}. 
%
It has been verified that our code 
reproduces the solution of ref.~\citenum{wedin:09} 
with high accuracy (choosing $\alpha h=0.85$ leads to
$\min(Re_b)=596.1191$ and $c/u_{max}=0.4780$ when using $N_x=4$,
$N_y=N_z=26$).   
%
\begin{figure}
  \begin{center}
    \begin{minipage}{6ex}
      $\displaystyle\lambda^+_{min}$
    \end{minipage}
    \begin{minipage}{.75\linewidth}
      \includegraphics[width=\linewidth]{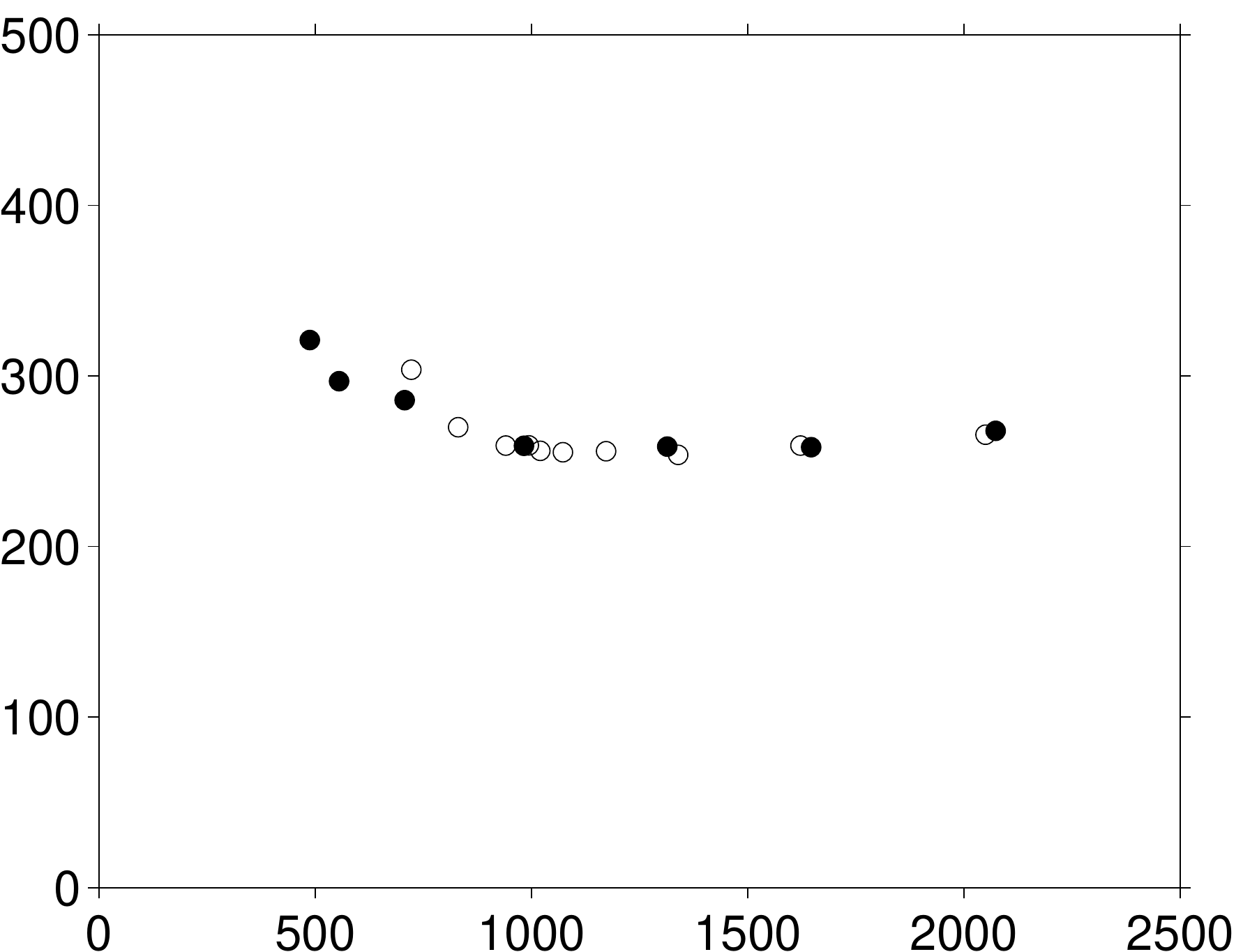}\\
      \centerline{$Re_{b}$}
    \end{minipage}
  \end{center}
  \caption{%
    Minimum wavelength (in wall units) of the travelling-waves given
    as a function of the bulk Reynolds number. The open symbols
    correspond to a truncation level of $2\times30\times30$, the
    filled ones to $4\times20\times20$. 
  }
  \label{fig-lambda_plus-min}
\end{figure}
\begin{figure}[b]
  \begin{center}
    \begin{minipage}{6ex}
      $\displaystyle\frac{E_{3D}^{1/2}}{u_b}$
    \end{minipage}
    \begin{minipage}{.75\linewidth}
      \includegraphics[width=\linewidth]{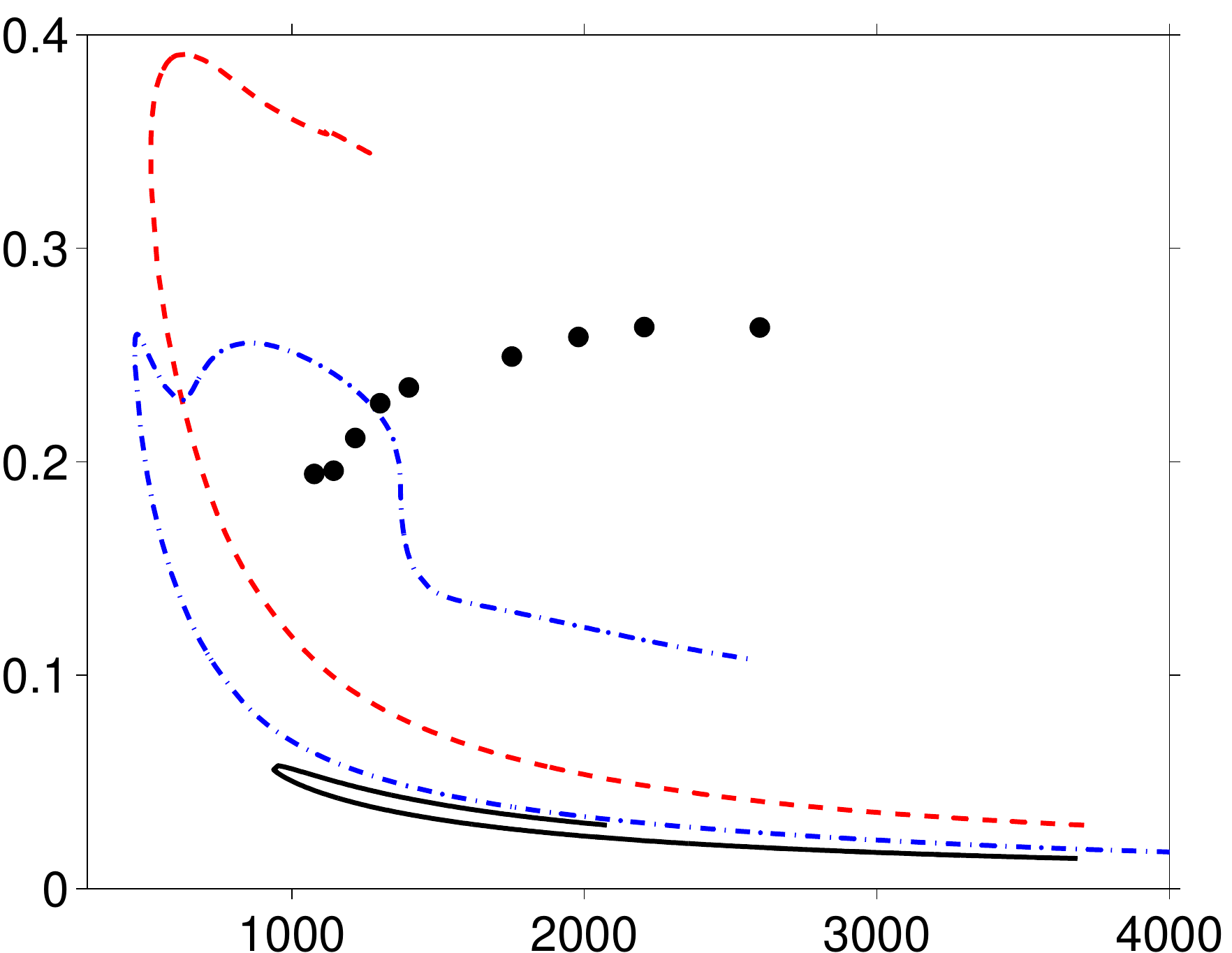}\\
      \centerline{$Re_{b}$}
    \end{minipage}
  \end{center}
  \caption{%
    Three-dimensional perturbation velocity, normalized by the bulk
    velocity, as a function of the bulk Reynolds number. 
    The lines connect solutions
    at constant streamwise wavenumber: $\alpha h=0.6$ ({\color{red}dashed}),  
    $\alpha h=1$ ({\color{blue}chain-dotted}), $\alpha h=1.5547$
    ({\color{black}solid}). 
    The symbols ($\bullet$) are for turbulent flow\cite{uhlmann:07a}. 
  }
  \label{fig-e3d}
\end{figure}
Furthermore, numerical convergence of the solutions presented in the
following has been thoroughly investigated.
Table~\ref{tab-convergence} shows the convergence with respect to the
number of streamwise Fourier modes as well as the cross-stream
polynomials for two particular parameter points. 
It was found that the truncation level required for resolving the
present solution family varies considerably in parameter space. 
In particular, resolving the upper branch solution at small
wavenumbers imposes the most stringent requirements (as
previously observed in Couette flow\cite{jimenez:05}). 
This point is illustrated in figure~\ref{fig-cf-conv1}, where solution
curves at two different streamwise wavenumbers are depicted as
computed with different truncation levels. 
The plotted quantity is the perturbation energy 
which excludes the streamwise-constant modes, i.e.\ 
$E_{3D}=\sum_{n\neq0}\int\int
|\hat{\mathbf{u}}_n|^2\mbox{d}y\mbox{d}z$ where $\hat{\mathbf{u}}_n$
is the $n$th Fourier coefficient of velocity. 
The figure shows that $N_x=2$ suffices for $\alpha h=1.5547$, while
the smaller wavenumber value $\alpha h=0.6$ requires a considerably
higher number of streamwise modes. 
In particular, we have used a streamwise truncation level of up to
$N_x=20$ on the upper branch for $\alpha h=0.6$. With a resolution 
of $N_x=20$ and $N_y=N_z=20$ we obtain results which are within one
percent of the results obtained for $N_x=14$ and $N_y=N_z=20$ up to
$Re=6565$ (corresponding to $Re_b=1247.2$). Beyond that point,
results obtained with these two truncation levels exhibit
substantially larger differences, and our finest solution will
therefore only be discussed up to $Re_b=1247.2$ in the
following. These tough resolution requirements stem from 
the fact that the flow field on the upper
branch at smaller wavenumbers develops very strong streamwise
gradients as can be seen in figure~\ref{fig-field} below.

\begin{figure}
  \begin{center}
    \begin{minipage}{5ex}
      $\displaystyle c_f$
    \end{minipage}
    \begin{minipage}{.75\linewidth}
      \includegraphics[width=\linewidth]{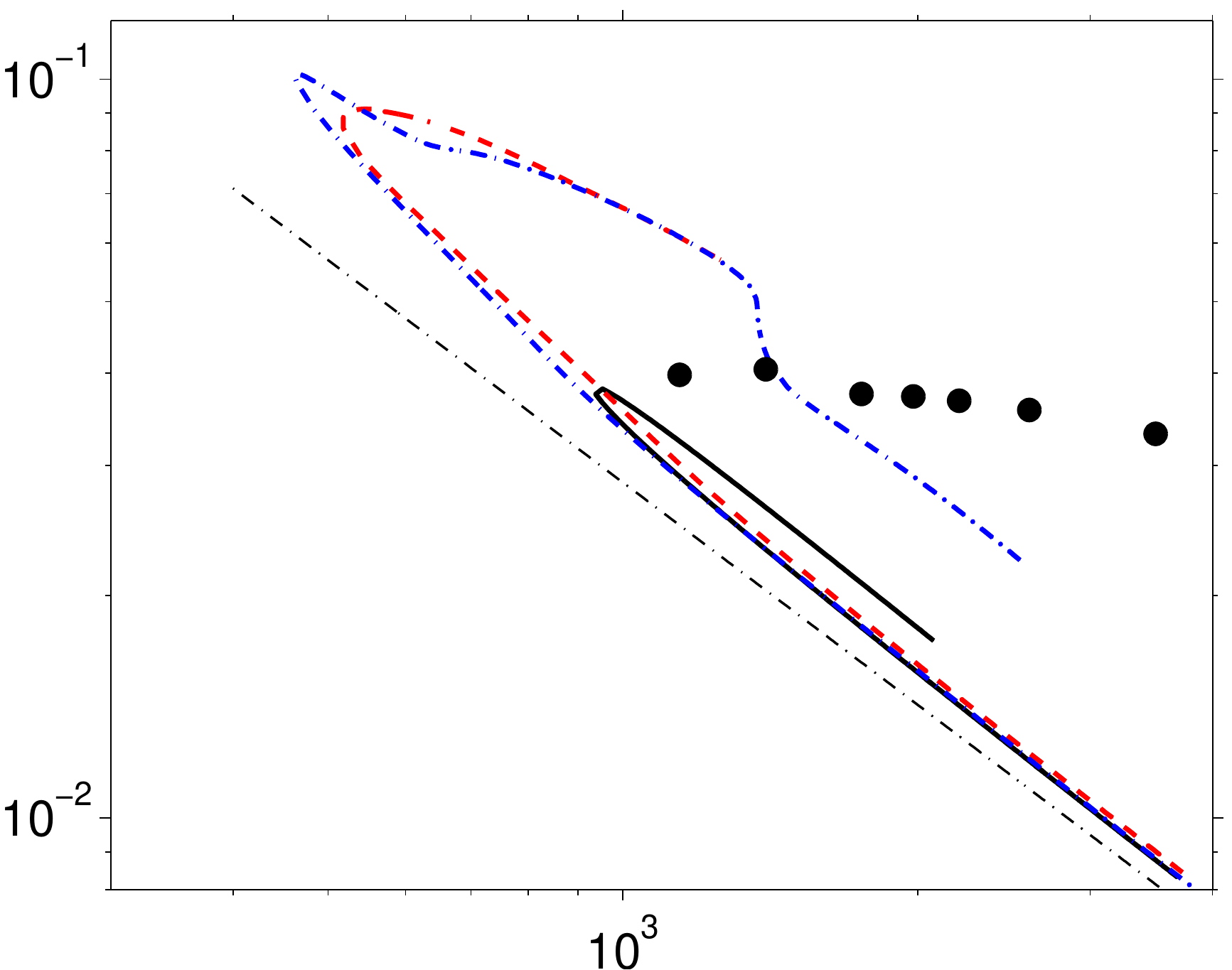}\\
      \centerline{$Re_b$}
    \end{minipage}
  \end{center}
  \caption{%
    Wall friction factor $c_f=8u_\tau^2/u_b^2$ as a function of
    $Re_b$ in logarithmic scale ($u_\tau$ being the friction velocity).
    Line styles and symbols as defined in figure~\ref{fig-e3d}. 
    The additional straight chain-dotted line indicates laminar friction. 
  }
  \label{fig-cf-reb}
\end{figure}
In the absence of a ``natural'' primary bifurcation point from which a
reasonable initial guess can be constructed, we resort to the 
homotopy method
proposed by Waleffe \cite{waleffe:03}
based on the so-called {\it
  self-sustaining process} of wall-bounded shear flows.
Here streamwise vortices are artificially added
to the base flow and forced against viscous decay, leading to streaks,
which are in turn linearly unstable, feeding back into the original
vortices. The non-linear solution is then continued back to the
original problem, i.e.\ the unforced flow. 
As suggested by Waleffe\cite{waleffe:03} we select the streamwise
rolls amongst the least decaying eigenmodes of the Stokes operator in
the cross-sectional plane. In the following we will focus upon results
obtained using the fourth eigenmode\cite{leriche:04} which exhibits a
parity of type III in the nomenclature of ref.~\citenum{tatsumi:90}
(cf.\ table~\ref{tab-parities}).
In fact, the non-linear equations admit solutions which are such that
all even streamwise Fourier modes verify the parity type III and all
odd Fourier modes simultaneously verify one of the parity types I,
II, III or IV. 
This property allows us to reduce the system size while choosing any
linear streak instability mode in conjunction with rolls 
of parity III. The following results were obtained 
when starting at a primary bifurcation point of a linear perturbation
with parity II. 
%
\begin{figure*}[hb!]
  \begin{minipage}{.35\linewidth}
    \includegraphics[width=\linewidth]
    {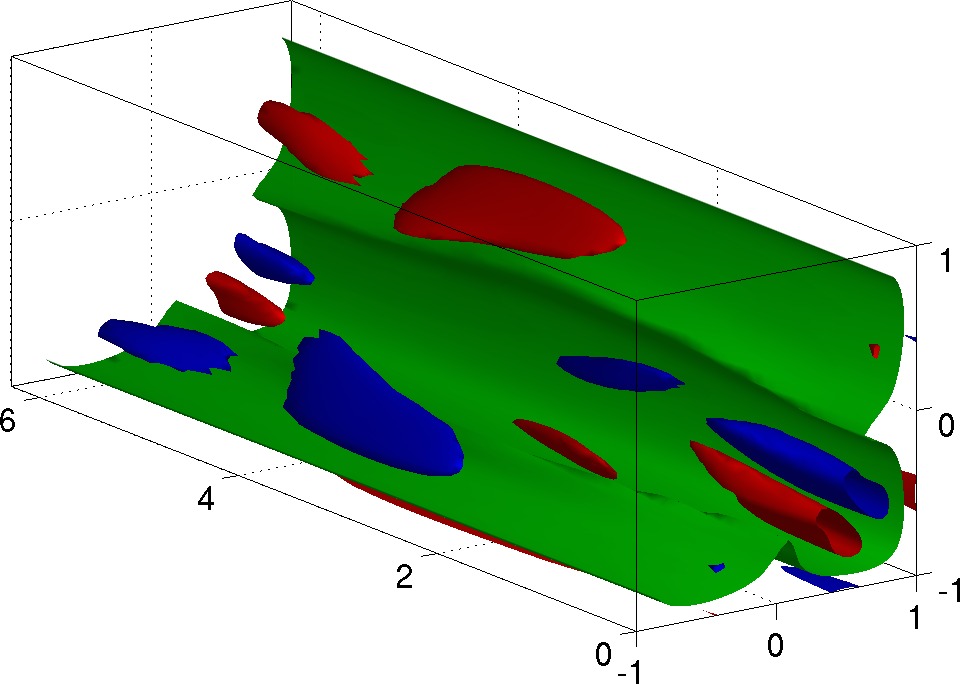}
    \hspace*{-.8\linewidth}\raisebox{.1\linewidth}{
    $x/h$}
    \hspace*{.3\linewidth}\raisebox{-.05\linewidth}{
    $z/h$}
  \end{minipage}
  \hfill
  \begin{minipage}{2ex}
    $\displaystyle\frac{y}{h}$
  \end{minipage}
  \begin{minipage}{.25\linewidth}
    \includegraphics[width=\linewidth]
    {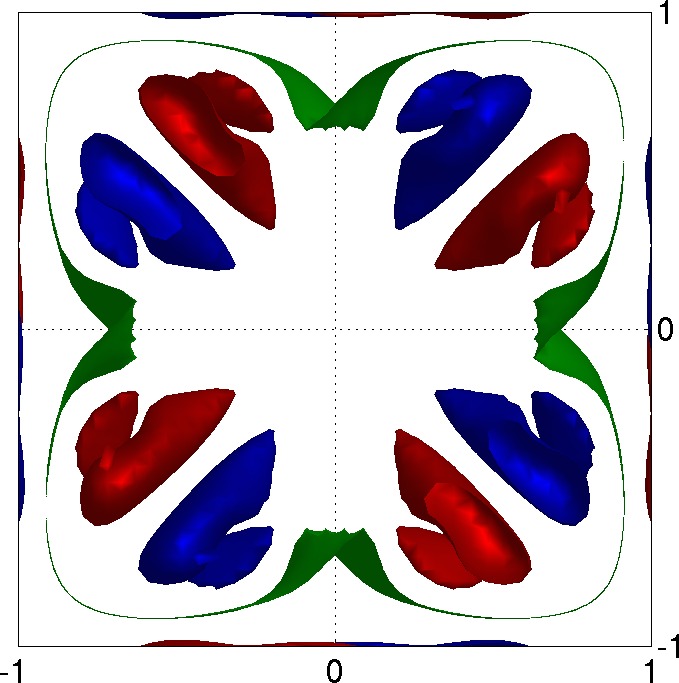}
  \end{minipage}
  \hfill
  \begin{minipage}{2ex}
    $\displaystyle\frac{y}{h}$
  \end{minipage}
  \begin{minipage}{.25\linewidth}
    \includegraphics[width=\linewidth]
    {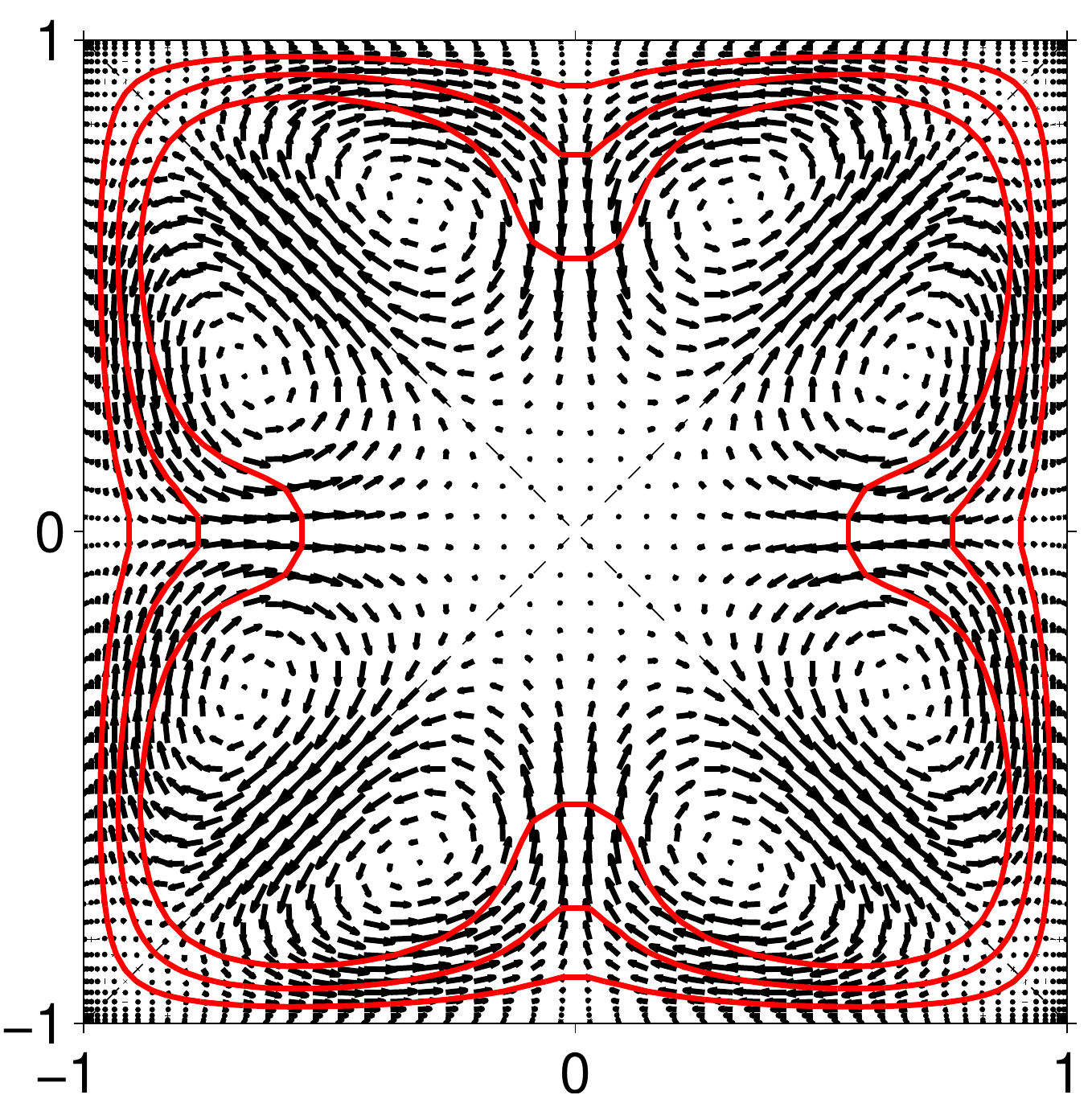}
  \end{minipage}
  \\[1ex]
  \begin{minipage}{.35\linewidth}
    \includegraphics[width=\linewidth]
    {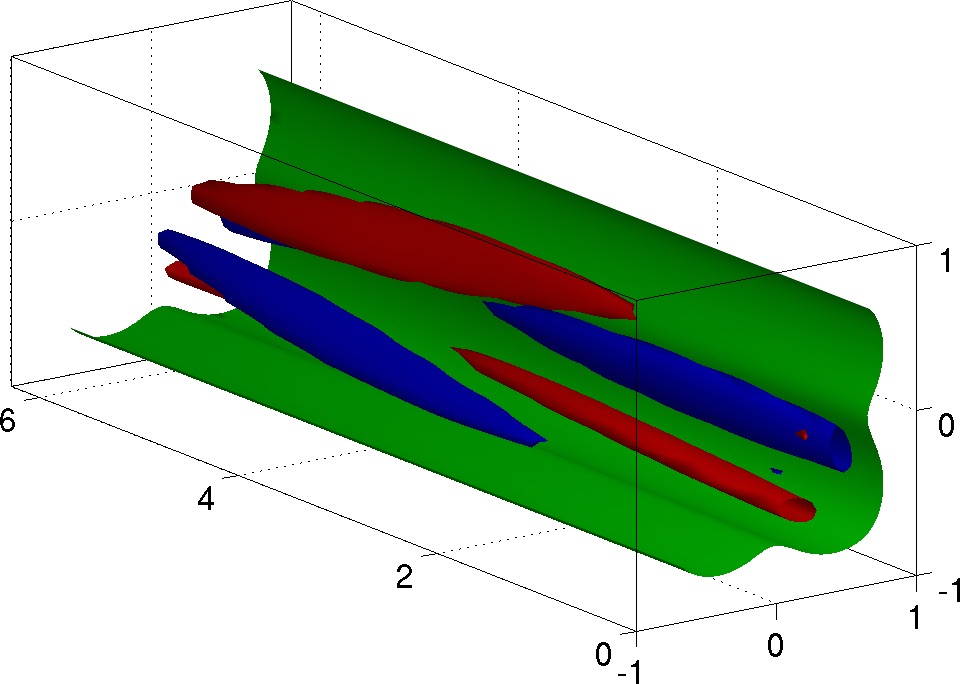}
    \hspace*{-.8\linewidth}\raisebox{.1\linewidth}{
    $x/h$}
    \hspace*{.3\linewidth}\raisebox{-.05\linewidth}{
    $z/h$}
  \end{minipage}
  \hfill
  \begin{minipage}{2ex}
    $\displaystyle\frac{y}{h}$
  \end{minipage}
  \begin{minipage}{.25\linewidth}
    \includegraphics[width=\linewidth]
    {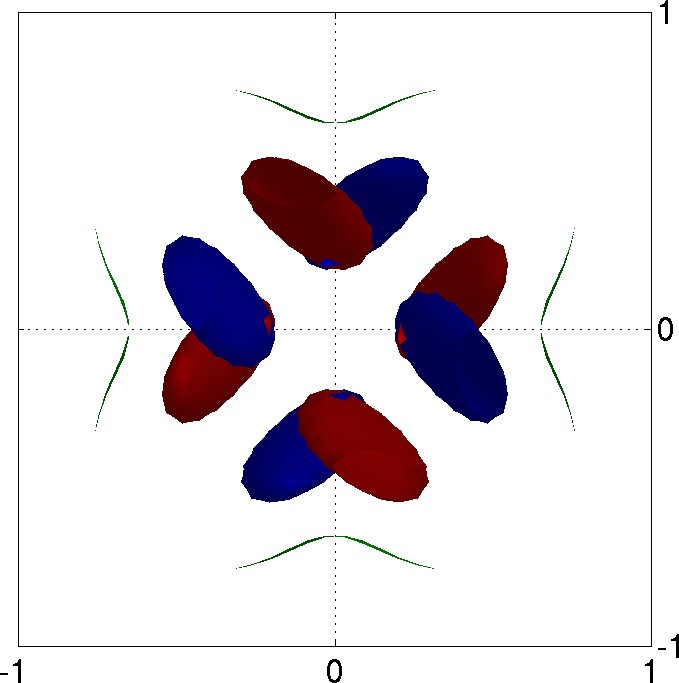}
    \\
    \centerline{$z/h$}
  \end{minipage}
  \hfill
  \begin{minipage}{2ex}
    $\displaystyle\frac{y}{h}$
  \end{minipage}
    \begin{minipage}{.25\linewidth}
    \includegraphics[width=\linewidth]
    {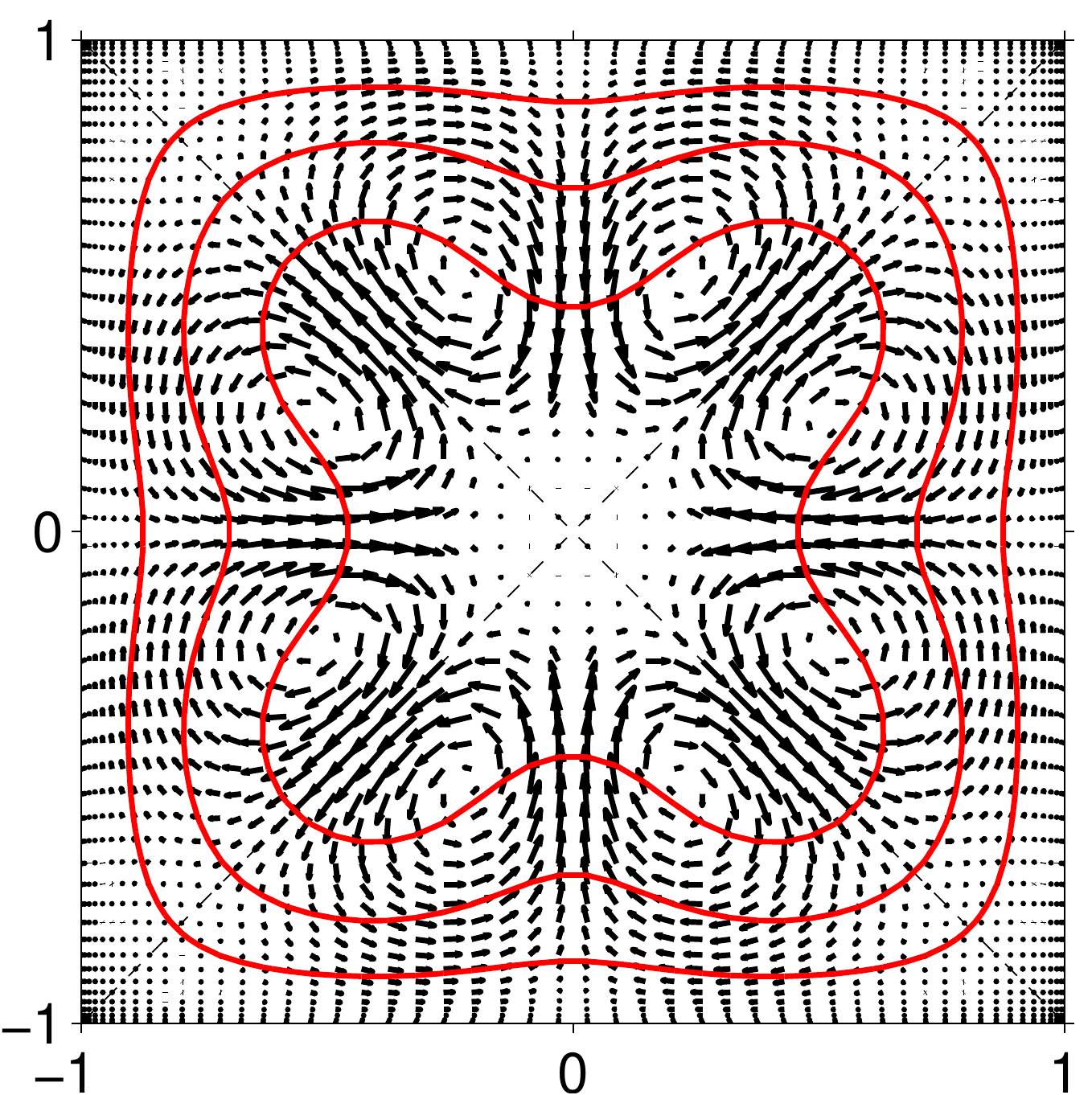}
    \\
    \centerline{$z/h$}
  \end{minipage}
  \caption{%
    Top row: the upper branch solution at $Re_b=1404.1$; 
    bottom row: the lower branch solution at $Re_b=1398.3$; both at
    $\alpha h=1$. 
    The graphs in the left and center column show surfaces of constant
    values of the total streamwise velocity $u=0.55\max(u)$ (sheet-like
    structure near the walls, green color) 
    and of the streamwise vorticity at $\pm0.65$ ($\pm0.55$) times its maximum
    absolute value in the case of the upper (lower) branch solution
    (tubular structures, blue/red color). 
    This value corresponds to $\omega_x^+=\pm0.0655$ ($\omega_x^+=\pm0.040$).
    In the left column the surfaces are only shown on one side of a
    diagonal of the duct cross-section.
    The right column depicts the corresponding mean flow,
    the primary flow $\langle u\rangle$ being visualized by contourlines at
    $\{0.25,0.5,0.75\}$ times the maximum value, and the secondary
    flow $(\langle v\rangle,\langle w\rangle)$ in form of arrows. 
    Truncation level: $N_x=20$, $N_y=N_z=26$.
  }
  \label{fig-field}
\end{figure*}
\section{Results}\label{sec-results}
The critical point (minimum value of the bulk Reynolds number) is
found at $\min(Re_b)=471$ for $\alpha h=0.927$ 
(determined at a truncation level of $N_x=4$, $N_y=N_z=26$); this
corresponds to a value of the friction-velocity based Reynolds number
of $Re_\tau=51.6$.
It should be recalled that self-sustained turbulence in square duct
flow exists above $Re_b\approx1070$\cite{uhlmann:07a}
($Re_\tau\approx80$);  
localized turbulent ``puffs'' have been found to subsist above
$Re_b\approx800$ (unpublished results). 
The region in parameter space which is occupied by this
travelling-wave family has an upper bound in wavenumber $\alpha$
which increases with the Reynolds number; when normalized in wall
units 
the corresponding short wavelength limit $\lambda^+_{min}$ is
relatively constant with a value between 250 and 300 over a
considerable range of $Re_b$ (figure~\ref{fig-lambda_plus-min}). 
This shortest streamwise wavelength at which the travelling-waves
appear is not very far from the length of the minimal flow unit for
self-sustained turbulence which measures roughly
190 wall units in the square duct\cite{uhlmann:07a} and 250--300 in
plane channel flow\cite{jimenez:91}.
It should be noted that a precise determination of the {\it upper}
limit of the streamwise wavelength as function of Reynolds number is
beyond the scope of the present paper. 

Figure~\ref{fig-e3d} characterizes the solution in terms of the
three-dimensional perturbation energy. Here and in the following we
will focus upon three specific values of the streamwise wavenumber,
namely $\alpha h=\{0.6,1.0,1.5547\}$.  
It can be seen that the quantity $E_{3D}^{1/2}/u_b$ varies
significantly with the streamwise wavenumber (especially on the upper
branch), the highest values being observed for the smallest of the
three wavenumbers ($\alpha h=0.6$).  
The figure also includes the time-averaged counterpart of the
three-dimensional perturbation energy for turbulent
flow (obtained through DNS\cite{uhlmann:07a}).
At low Reynolds number the present solutions with wavenumber $\alpha
h=\{0.6,1\}$ are more energetic than the average turbulent flow, while
the opposite is true for $\alpha h=1$ at higher Reynolds number
$Re_b\gtrsim1370$; unfortunately, 
due to the above mentioned high resolution requirements, 
it was not possible to further continue the
curve for $\alpha h=0.6$ in order to judge whether a similar drop-off
occurs at that wavenumber.  

Let us now turn to the friction factor. 
Figure~\ref{fig-cf-reb} demonstrates that the lower branch of the
present solution surface scales approximately as the laminar solution
(i.e.\ varying as $Re_b^{-1}$) at all wavenumbers.
The upper branch exhibits significantly higher friction values at
smaller wavenumbers. On the upper branch curves corresponding to
$\alpha h=0.6$ and $\alpha h=1$ the average turbulent friction values
are exceeded for small Reynolds numbers; again, cross-over between the
travelling-waves at $\alpha h=1$ and the turbulent data takes place at
$Re_b\approx1370$, beyond which the upper branch solution scales
approximately as $Re_b^{-1}$. 
Conversely, for the larger wavenumber ($\alpha
h=1.5547$) the upper branch has a similar slope to the lower branch at
all Reynolds numbers and does not correspond to significantly
increased friction with respect to the laminar base flow. 

In figure~\ref{fig-field} solutions on the upper and lower branch at
the same wavelength ($\alpha h=1$) and similar bulk
Reynolds number ($Re_b\approx1400$) are
depicted by means of isosurfaces of streamwise vorticity $\omega_x$
and total streamwise velocity $u$. The former isosurfaces take the 
shape of streamwise elongated tubes which are slightly inclined with
respect to the wall plane. Surfaces of the streamwise velocity exhibit
a single corrugation per wall, which corresponds to a low-speed
streak. The streamwise vortices are in streamwise alternating
(staggered) arrangement with respect to the nearest streak.  
When focusing upon structures in the vicinity of a single wall, it
becomes clear that the present travelling-waves are similar to exact
coherent structures found in plane wall-bounded shear
flows\cite{waleffe:03}.  
Furthermore, the staggered vortex/low-speed streak configuration
described above is at the center of a multitude of observations in
the near-wall region of turbulent flow (e.g.\ DNS of Jeong et
al.\cite{jeong:97}) as well as in time-periodic
solutions\cite{kawahara:01b}.    

The lower branch solution in figure~\ref{fig-field} exhibits a very
mild streamwise dependency (i.e. the streaks are nearly straight), as
well as very smoothly shaped, large-scale streamwise vortices. This is
consistent with the fact that the shape of the lower branch solution
changes very little with Reynolds number.
The upper branch, on the other hand, is characterized by very 
localized vortices and relatively sharp bending of the streaks. The
high intensity of the vortices in the upper branch
solution as well as their proximity to the walls
is manifested through the existence of patches with strong mirror
vorticity.  

Figure~\ref{fig-field} also shows the streamwise-averaged flow
corresponding to the upper/lower branch solutions. 
A secondary flow pattern with eight vortices, very much alike the one
exhibited on average by turbulent duct flow, can be observed. 
The resulting deformation of the primary flow contours leads
to a variation of wall-shear along the edges (not shown), which is
characterized by two local maxima near the corners and a local
minimum at the bisector -- as found in turbulent
flow\cite{pinelli:09a} at Reynolds numbers below $Re_b\approx2000$ . 
Figure~\ref{fig-mean-flow-dns} shows the statistically
  averaged mean flow from DNS\cite{pinelli:09a} at $Re_b=1400$, 
  facilitating a direct comparison with the present results of
  figure~\ref{fig-field} at practically the same Reynolds number. A
  strong similarity of primary flow contours and secondary flow
  pattern between DNS data and the upper branch travelling-wave can
  indeed be observed.   

\begin{figure}
  \begin{center}
    \begin{minipage}{5ex}
      $\displaystyle\frac{E_{vw}^{1/2}}{u_b}$
    \end{minipage}
    \begin{minipage}{.75\linewidth}
      \includegraphics[width=\linewidth]{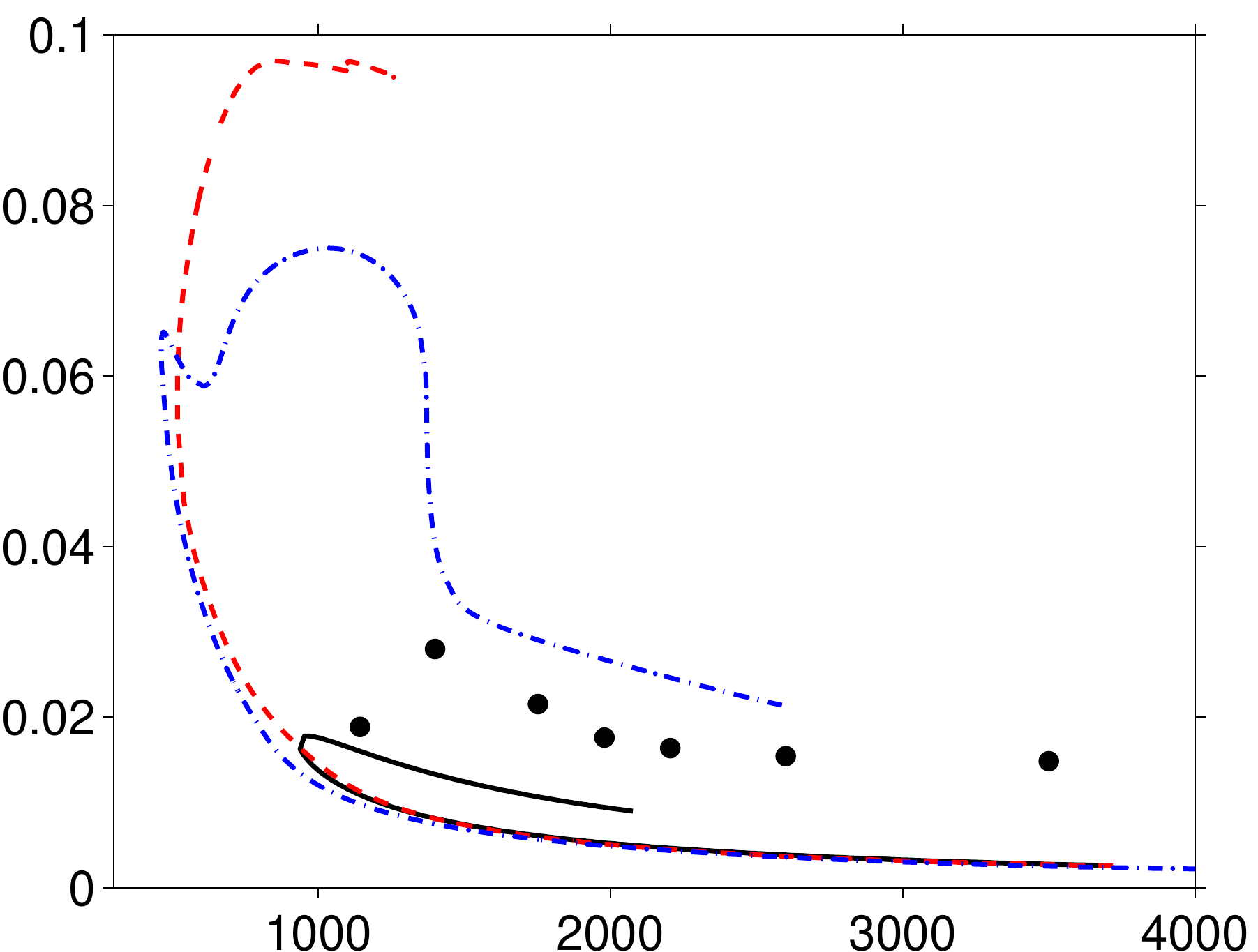}\\
      \centerline{$Re_b$}
    \end{minipage}
   \end{center}
  \caption{%
    The secondary flow intensity of the present solution family,
    normalized by the bulk velocity and shown as a function of
    $Re_b$. 
    Line styles and symbols as defined in figure~\ref{fig-e3d}.
  }
  \label{fig-enervw-reb}
\end{figure}

In order to quantify the amplitude of the streamwise
averaged secondary motion the following energy norm can be defined: 
$E_{vw}=\int\int
(|\hat{v}_0|^2+|\hat{w}_0|^2)\mbox{d}y\mbox{d}z$.  
The square-root of this quantity, normalized by the bulk velocity, is
shown in figure~\ref{fig-enervw-reb}. 
Analogously to the perturbation energy (cf.\ figure~\ref{fig-e3d}),
the present travelling-waves (at small wavenumbers) by far surpass
the average secondary flow intensity of turbulence at lower Reynolds
numbers. Again a steep drop occurs for $\alpha h=1$ around
$Re_b\approx1370$, after which the travelling-waves and turbulence
exhibit secondary flow of similar magnitude (to within 50\% of each
other). 
While the turbulent data for $E_{vw}^{1/2}/u_b$ appears to approach a
constant value at higher Reynolds numbers, no scaling of the
present upper-branch solutions can be deduced from our data. 
On the other hand, the curves for the lower branch practically
collapse at all wavenumbers and vary approximately as
$E_{vw}^{1/2}/u_b\sim Re_b^{-1}$, 
implying a scaling with the viscous velocity, i.e.\
$E_{vw}^{1/2}\sim\nu/h$; 
this scaling has previously been observed in Couette
flow\cite{wang:07} and pipe flow\cite{viswanath:08}. 

\begin{figure}
   \begin{center}
     \begin{minipage}{5ex}
       $\displaystyle\frac{c}{u_b}$
     \end{minipage}
     \begin{minipage}{.75\linewidth}
       \includegraphics[width=\linewidth]{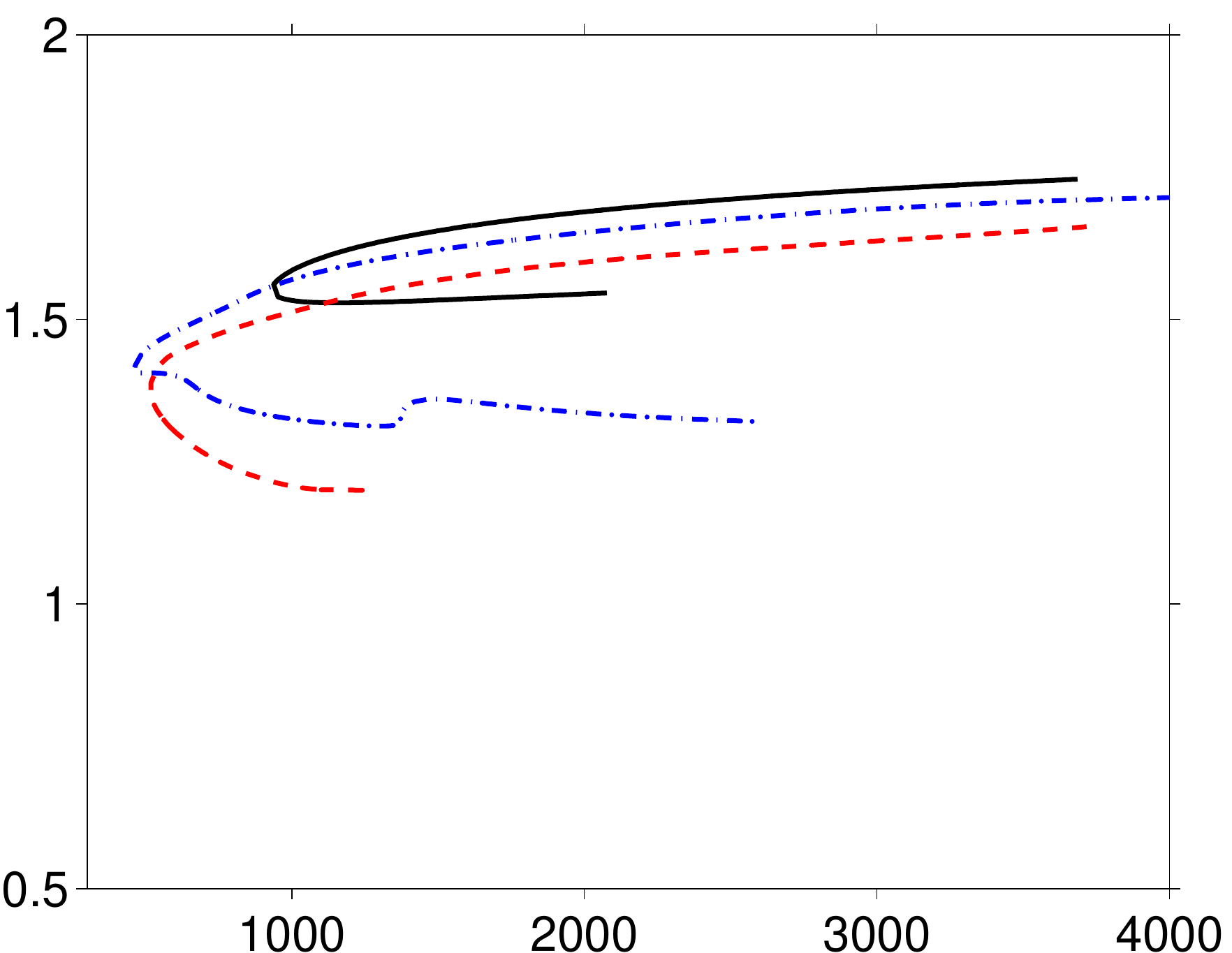}\\
       \centerline{$Re_b$}
     \end{minipage}
   \end{center}
   \caption{%
       The phase velocity normalized by the bulk velocity. 
       Line styles as defined in figure~\ref{fig-e3d}.
   }
   \label{fig-cphase-reb}
 \end{figure}

The phase velocity 
(shown in figure~\ref{fig-cphase-reb}) 
corresponding to the present
lower-branch solutions tends towards values in the range of
$c/u_b=1.66-1.75$ (depending on the wavenumber), which is
comparable to the asymptotic value $1.76$ of  
the lower-branch solution with mirror symmetry in pipe
flow\cite{viswanath:08}. 
On the upper branch and for small wavenumbers, the values for the
phase speed of the present solutions are in the range of
$c/u_b=1.2-1.4$. In terms of the friction velocity, the lowest
phase-speed $c^+_{min}\approx12$ is obtained near the turning point,
i.e.\ the junction of lower and upper branches (for both wavenumbers
$0.6$ and $1$). This value is comparable to the propagation speed of
near-wall coherent structures in turbulent channel flow, which is
equal to approximately $10$ times the friction velocity\cite{kim:93c}.   
%
\section{Conclusion}\label{sec-conclusion}
The present non-linear equilibrium solutions to the Navier-Stokes
equations are characterized by coherent structures (staggered
streamwise vortices and wavy streaks) commonly found in wall-bounded
shear flows. Their streamwise average also corresponds to the pattern
of the time-averaged velocity field observed in (laboratory and
numerical) experiments of turbulent square duct flow. Therefore, they
demonstrate (for the first time theoretically) that near-wall coherent
structures are indeed 
capable of generating physically correct
secondary motion in duct flow (as previously
conjectured\cite{uhlmann:07a,pinelli:09a}). 
Furthermore, we have shown that the present family of travelling-waves
yields integral flow quantities 
which are generally comparable to statistical data of turbulent flow.
This implies that these equilibrium solutions are potentially relevant
to the turbulent state. 
In particular, 
it was shown that the upper branch solution at an intermediate streamwise
wavenumber ($\alpha h=1$) 
yields secondary flow intensities which are
  comparable to turbulence data for $Re_b\geq1370$; furthermore, the
  upper-branch solution 
  changes from a highly energetic and high wall friction state at lower
  Reynolds numbers to a low-intensity state for higher $Re_b$,
  crossing the statistical energy and wall friction level of turbulence
  around $Re_b\approx1370$.  
Interestingly, this drop in perturbation energy and wall friction 
coincides with the Reynolds number range where turbulent duct flow
exhibits a transition in terms of the instantaneous flow pattern: 
below $Re_b\approx1250$ the predominant state of turbulent flow is
characterized by a vortex pair and a single low-speed streak adjacent
to one pair of opposite walls (the so-called 4-vortex state), whereas
this feature gradually disappears with increasing Reynolds number,
giving way to an 8-vortex state (vortex pair/streak present near all
four walls simultaneously)\cite{uhlmann:07a}. 
Therefore, it can be expected that the present solution family is 
pertinent to turbulence-induced mean secondary motion in square
ducts at $Re_b\gtrsim1370$. 
Furthermore, DNS shows that the average number of streaks which are
simultaneously present along one edge of the cross-section increases
with the Reynolds number\cite{pinelli:09a}, consistent with a
scaling of the spanwise streak-spacing in wall units. This property
implies that additional travelling-waves with a distinct vortex
pattern (16, 24, 32,... vortices in total, i.e.\ 5, 7, 9,... streaks
per edge) are expected to come into play progressively as the Reynolds
number is increased.  
As a consequence, we expect our present family of exact coherent
structures to be consistent with actual buffer-layer structures in
turbulent flow only around $Re_b\approx1370$. 

It should be mentioned that other families of travelling-waves which
are also consistent with the coherent buffer-layer structures and the
average secondary flow pattern encountered in square duct turbulence
might very well exist. 
However, the central point of the present communication is the fact
that such a non-linear solution does indeed exist at all.

In order to determine which role the present equilibrium solutions play in
the dynamics of the initial value problem, 
the stability characteristics of the travelling-waves should first be
investigated. Then    
actual turbulent phase-space trajectories need to be mapped out, and
their relation to the travelling waves (fixed points) needs to be
established (as e.g.\ performed for Couette flow in
ref.~\citenum{gibson:08}).  
This analysis will be the object of a future study. 
\section*{Acknowledgments}
\begin{figure}
  \begin{center}
    \begin{minipage}{2ex}
      $\displaystyle\frac{y}{h}$
    \end{minipage}
    \begin{minipage}{.5\linewidth}
      \includegraphics[width=\linewidth]
      {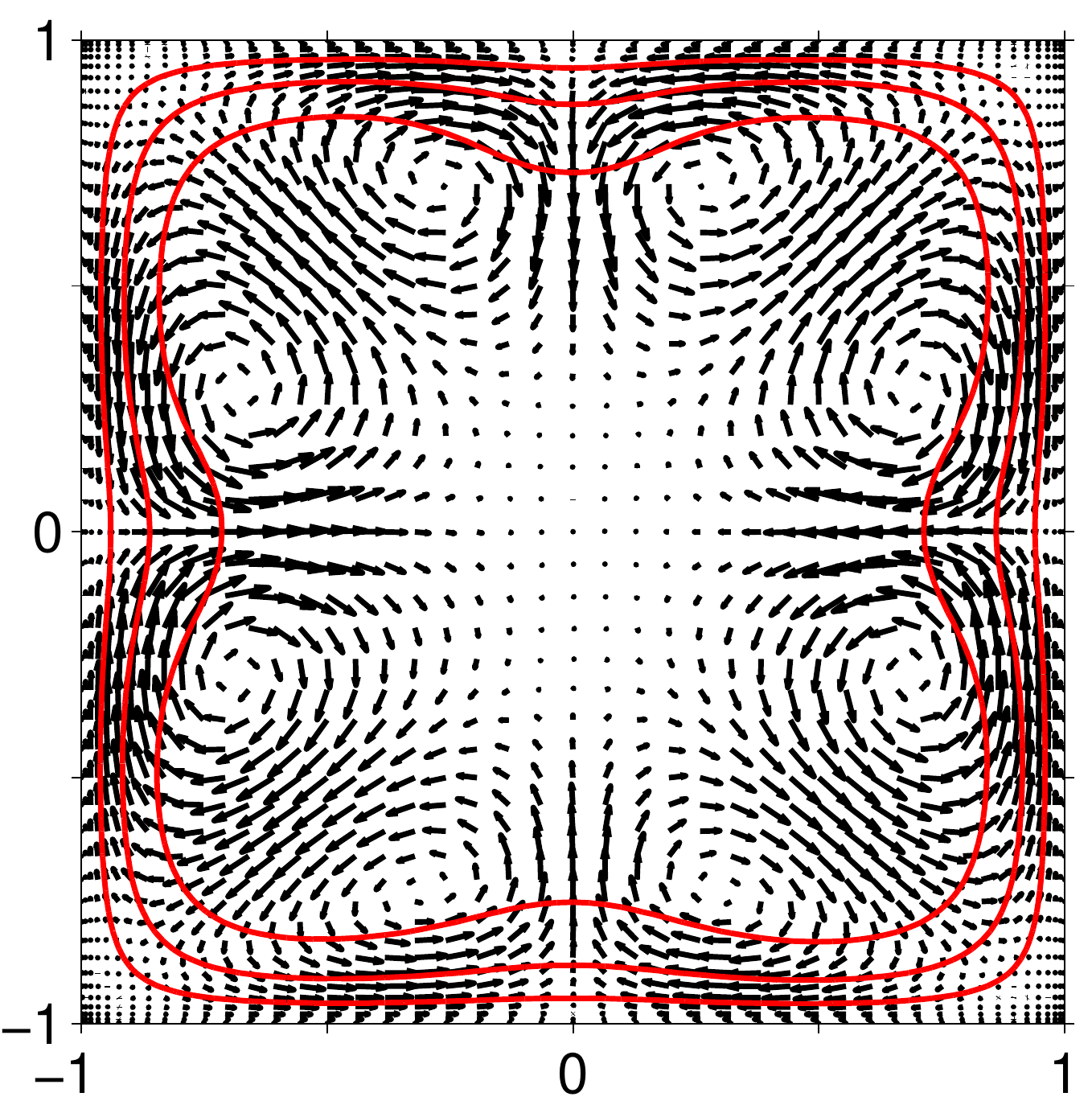}
      \\
      \centerline{$z/h$}
    \end{minipage}
  \end{center}
  \caption{%
    Statistical mean flow from DNS\cite{pinelli:09a} at
    $Re_b=1400$. Contour levels as in
    figure~\ref{fig-field} (right column).
  }
  \label{fig-mean-flow-dns}
\end{figure}

The authors are grateful to A.\ Sekimoto for providing the DNS data
visualized in figure~\ref{fig-mean-flow-dns}.
\bibliographystyle{unsrt}

\end{document}